\documentclass[12pt,twoside]{article}
\usepackage{mathpazo} 
\usepackage[mathscr]{eucal}
\usepackage{amsmath,amsfonts,amssymb,amsthm}
\usepackage{times}
\usepackage{url}
\usepackage{hyperref}

\def\d_Vphi{\text{d}_V\hspace{-0.06em}\phi}
\def\d_Vphibar{\text{d}_V\hspace{-0.06em}\bar\phi}
\def\d_Vxi{\text{d}_V\hspace{-0.06em}\xi}

\def\be{\begin{eqnarray}}
\def\ee{\end{eqnarray}}
\def\beann{\begin{eqnarray*}}
\def\eeann{\end{eqnarray*}}
\def\beq{\begin{equation}}
\def\eeq{\end{equation}}
\def\ba{\begin{array}}
\def\ea{\end{array}}
\def\ben{\begin{enumerate}}
\def\een{\end{enumerate}}
\def\bea{\begin{eqnarray}}
\def\eea{\end{eqnarray}}

\def\5{\bar }
\def\6{\partial }
\def\7{\hat }
\def\4{\tilde }
\advance\textheight\topskip
\renewcommand{\tilde}{\widetilde}
\renewcommand{\hat}{\widehat}

\newcommand{\dd}{\partial}
\renewcommand{\d}{\partial}

\renewcommand{\geq}{\,{\geqslant}\,}
\renewcommand{\leq}{\,{\leqslant}\,}

\newcommand{\binner}[2]{%
  {\langle}\kern-4.15pt{\langle}#1{,}\,#2{\rangle}\kern-4.15pt{\rangle}}

\newcommand{\half}{\frac{1}{2}}

\newcommand{\ffrac}[2]{\raisebox{.5pt}%
  {\footnotesize$\displaystyle\frac{#1}{#2}$}\kern1pt}

\newcommand{\dover}[2]{\ffrac{\dd #1}{\dd #2}}

\newcommand{\ddl}[2]{\ffrac{\dd #1}{\dd #2}}

\newcommand{\vddl}[2]{{\ffrac{\delta #1}{\delta #2}}}

\def\cJ{\mathcal{J}}
\def\cK{\mathcal{K}}
\def\cL{\mathcal{L}}
\def\cM{\mathcal{M}}
\def\cN{\mathcal{N}}
\def\cO{\mathcal{O}}

\def\cY{\mathcal{Y}}

\numberwithin{equation}{section} \makeatletter
\@addtoreset{equation}{section}

\DeclareFontFamily{OT1}{rsfs}{} \DeclareFontShape{OT1}{rsfs}{m}{n}{
<-7> rsfs5 <7-10> rsfs7 <10-> rsfs10}{}
\DeclareMathAlphabet{\mycal}{OT1}{rsfs}{m}{n}
\def\scri{{\mycal I}}%

\newcommand*\xbar[1]{%
  \hbox{%
    \vbox{%
      \hrule height 0.5pt 
      \kern0.3ex
      \hbox{%
        \kern-0.0em
        \ensuremath{#1}%
        \kern-0.0em
      }%
    }%
  }%
} 

\begin{document}

\author{Glenn Barnich and C\'edric Troessaert}

\title{Comments on holographic current algebras and asymptotically
  flat four dimensional spacetimes at null infinity}

\date{}

\def\mytitle{Comments on holographic current algebras and
  asymptotically flat four dimensional spacetimes at null infinity}

\pagestyle{myheadings} \markboth{\textsc{\small G.~Barnich and
    C.~Troessaert}}{%
  \textsc{\small Holographic current algebras and BMS4 }}
\addtolength{\headsep}{4pt}


\begin{centering}

  \vspace{1cm}

  \textbf{\Large{\mytitle}}


  \vspace{1.5cm}

  {\large Glenn Barnich$^{a}$}

\vspace{.5cm}

\begin{minipage}{.9\textwidth}\small \it  \begin{center}
   Physique Th\'eorique et Math\'ematique \\ Universit\'e Libre de
   Bruxelles and International Solvay Institutes \\ Campus
   Plaine C.P. 231, B-1050 Bruxelles, Belgium
 \end{center}
\end{minipage}

\vspace{.5cm}

{\large C\'edric Troessaert$^{b}$}

\vspace{.5cm}

\begin{minipage}{.9\textwidth}\small \it  \begin{center}
   Centro de Estudios Cient\'{\i}ficos (CECs) \\ 
Casilla 1469, Valdivia, Chile
 \end{center}
\end{minipage}

\end{centering}

\vspace{1cm}

\begin{center}
  \begin{minipage}{.9\textwidth}
    \textsc{Abstract}.  We follow the spirit of a recent proposal to
    show that previous computations for asymptotically flat spacetimes
    in four dimensions at null infinity can be re-interpreted in terms
    of a well-defined holographic current algebra for the time
    component of the currents. The analysis is completed by the
    current algebra for the spatial components.
  \end{minipage}
\end{center}

\vfill

\noindent
\mbox{}
{\scriptsize$^a$Research Director of the Fund for Scientific
  Research-FNRS Belgium. E-mail: gbarnich@ulb.ac.be\\
  $^b$ Laurent Houart postdoctoral fellow. E-mail: troessaert@cecs.cl}

\thispagestyle{empty}
\newpage

\begin{small}
{\addtolength{\parskip}{-1.5pt}
 \tableofcontents}
\end{small}
\newpage

\section{Introduction}
\label{sec:introduction}

In recent work, Strominger \cite{strominger:2013,Strominger:2013lka}
has proposed to replace the computation of integrated surface charges
and their algebra associated to asymptotic symmetries in gravitational
or gauge theories by a local computation in terms of Ward identities
and current algebras.

A formal, path integral based, derivation of local Ward identities
involving Noether currents and their divergences (see
e.g.~\cite{DiFrancesco:1997nk}, sections 2.4 and 2.5) is essentially
equivalent to classical properties of these currents. In order to take
into account quantum effects, one has to use operator product
expansion techniques or use perturbative quantum field theory with due
care devoted to renormalization effects (see
e.g.~\cite{jackiw:1985}). 

In this note, we start with some brief comments on specific aspects of
the classical part of this construction, first for global and then for
gauge currents. It is the latter that are relevant in holographic
applications when performing computations on the bulk side of the
correspondence. As an illustration, the cases of three dimensional
asymptotically AdS spacetimes at spatial infinity \cite{Brown:1986nw}
and asymptotically flat spacetimes at null infinity
\cite{Barnich:2006avcorr,Barnich:2010eb,Barnich:2012aw} are
reconsidered with a special emphasis on the additional spatial current
components.

We then turn to asymptotically flat spacetimes at null infinity in
four dimensions. Re-interpreting previous results derived in
\cite{Barnich:2011mi} and translated to the Newman-Penrose formalism
in \cite{Barnich:2011ty} gives the algebra of the time component of
the currents. This is completed by working out the algebra of the
spatial components. The holographic current algebra is derived in a
unified way both for the standard, globally well defined asymptotic
symmetry algebra $\mathfrak{bms}_4^{\rm glob}$ involving Lorentz
transformations and supertranslation generators that can be expanded
into spherical harmonics as well as for its local version
$\mathfrak{bms}_4^{\rm loc}$ \cite{Barnich:2009se}.

The main benefit of the local formulation in terms of currents in this
context is that, for the local version of the asymptotic algebra,
there is no longer any problem with divergences related to poles as
there is no need to explicitly integrate the time component of the
currents over the sphere. Instead, one can now use contour integrals
on the unit or Riemann sphere if one so wishes.

In other words, in the standard approach the existence of
well-defined charges is taken as a criterion to reduce the asymptotic
symmetry algebra to $\mathfrak{bms}_4^{\rm glob}$, assuming of course
that the fields that enter the time components of the currents are
integrable on the sphere. Changing this criterion to the existence of
a well-defined local current algebra allows one to consistently deal
with $\mathfrak{bms}_4^{\rm loc}$.

\section{Classical current algebras}
\label{sec:class-curr-algebr}

\subsection{Global symmetries}
\label{sec:global-symmetries}

For an action principle that is invariant under global symmetry
transformations, there is a short-cut that allows one to determine the
Poisson or Dirac bracket algebra of the generators of these
transformations, without the need to go through the steps of the
Hamiltonian analysis. 

While the latter consists in first determining the brackets of the
fundamental canonical variables, then expressing the Noether charges
in these variables and finally evaluating their brackets, the short-cut
is well known (see e.g.~\cite{Dickey:1991xa,Barnich:1996mr}) and goes
as follows. Let $\cL=Ld^nx$ be the Lagrangian $n$ form of the
theory. Invariance means that $\delta_X \cL=d_H n_X$. Here,
$\delta_X\phi^i$ denotes the infinitesimal transformations of the
fields $\phi^i$ and $n_X$ an $n-1$ form. The differential $d_H=dx^\mu
\d_\mu$ involves the total derivative that takes into account the
space-time dependence of the fields, $\d_\mu=\dover{}{x^\mu}+\d_\mu
\phi^i\dover{}{\phi^i}+\dots$.

The Noether current $j_X=j^\mu_X (d^{n-1} x)_\mu$ associated to the
transformation $\delta_X$ satisfies
\begin{equation}
  \label{eq:1}
  X^i\vddl{\cL}{\phi^i}=d_H j_X,
\end{equation}
and can be chosen as $j_X=n_X-I^n_X(\cL)$, where
$I^n_X(\cL)=(X^i\dover{L}{\d_\mu\phi^i}+\dots) (d^{n-1}x_\mu)$.  The
Noether current is ambiguous, $j_X\sim j_X + d_H \eta_X + t_X$ where
$t_X$ is a Noether current that vanishes on-shell, while $\eta_X$ is
an $n-2$ form. By applying the symmetry transformation $-\delta_{X_2}$
to \eqref{eq:1} for a symmetry characterized by $X_1$, it is then
straightforward to show that
\begin{equation}
  \label{eq:2}
  -\delta_{X_2} j_{X_1}\sim j_{[X_1,X_2]} + K_{X_1,X_2}, 
\end{equation}
where the Lie bracket is determined through $[X_1,X_2]^i=\delta_{X_1}
X_2^i-(1\leftrightarrow 2)$, while the classical extension
$K_{X_1,X_2}$ belongs to $H^{n-1}(d_H)$ and may in general be field
dependent.

For instance, working out the classical current algebra of a chiral
Wess-Zumino-Witten model in this way is straightforward, while the
complete Hamiltonian analysis involves first and second class constraints
and is much more involved. 

\subsection{Gauge and asymptotic symmetries}
\label{sec:gauge-symmetries}

For asymptotic symmetries, which are a subset of the bulk gauge
symmetries, similar results can be shown under suitable
assumptions. Some elements of the general theory and more details can
be found for instance in
\cite{Regge:1974zd,Benguria:1976in,Abbott:1981ff,%
  Abbott:1982jh,Brown:1986nw,Brown:1986ed,Wald:1993nt,%
  Iyer:1994ys,Iyer:1995kg,Anderson:1996sc,Wald:1999wa,Barnich:2001jy,%
  Barnich:2003xg,Barnich:2007bf,Barnich:2010xq}.  

For a gauge transformation, $\delta _f\phi^i =
R^i_\alpha(f^\alpha)=R^i_\alpha f^\alpha+R^{i\mu}_\alpha \d_\mu
f^\alpha +\dots $ with possibly field dependent gauge parameters
$f^\alpha$, there is a weakly vanishing Noether current
$S_f=(R^{i\mu}_\alpha
f^\alpha\vddl{L}{\phi^i}+\dots)(d^{n-1}x)_\mu$. It is used to
construct a canonical representative for an $n-2$-form associated to
gauge symmetries through
\begin{equation}
k_f[\delta\phi]=I^{n-1}_{\delta\phi} S_f=\half \delta\phi^i \ddl{}{\d_\nu
  \phi^i}\ddl{}{dx^\nu} S_f +\dots \label{eq:12}. 
\end{equation}
Under standard assumptions, one can then show that, $d_H
k_{f_s}[\delta\phi_s]\approx 0$, when $f^\alpha_s$ satisfy
$R^i_\alpha(f^\alpha_s)\approx 0$, and when $\delta\phi_s$ satisfies the
linearized field equations. Here $\approx 0$ means for all solutions
of the Euler-Lagrange equations for $\phi$. The algebra of the forms
$k_f[\delta\phi]$ is obtained by applying a gauge transformation, and one
can show that
\begin{equation}
  \label{eq:5}
  -\delta_{f_2} k_{f_1} =k_{[f_1,f_2]} +{\rm more}, 
\end{equation}
where $[f_1,f_2]^\alpha$ involves the structure functions of the gauge
algebra. The expression for the additional terms can be worked out
quite generally, but what remains depends on the particular situation
that one considers. As in the global case, there will be trivial terms
proportional to the equations of motion, both for $\phi$ and
$\delta\phi$, and $d_H$ exact terms. The non-trivial terms include
possibly field-dependent central extensions.

In the asymptotic context, on which we will concentrate below, a
typical situation is to focus on a fixed hypersurface, say $r={\rm
  cte}\to\infty$, with prescribed asymptotic conditions on the
fields. What goes under the name of asymptotic symmetries is a
suitable sub-Lie algebroid of the Lie algebroid associated to gauge
symmetries. On the surface $r={\rm cte}\to\infty$, the $n-2$ form
$k_f=k^{[\mu\nu]}_f(d^{n-2}) x_{\mu\nu}$ becomes to an $(n-1)-1$
form. In a coordinate system $x^\mu=(u,r,y^A)$ for instance, the
components of the associated current for the lower dimensional theory
are given by $(k_f^{[ur]}, k^{[Ar]}_f)$. In the case where these
currents are integrable in solution space, $k^{[ur]}_f\approx \delta
J^u_f$, $k^{Ar}_f\approx\delta J^A_f$, their integrands $J^a_f$,
$x^a=(u,y^A)$, provide the global current algebra of the dual boundary
theory. The multiplicative normalization of $k_f$, and thus also of
$J_f$, is fixed through the action of the theory, whereas the
integration in solution space implies that the definition of the
integrands $J^a_f$ involves the choice of a background solution.

Even though the original computations for asymptotically anti-de
Sitter space-times \cite{Henneaux:1985tv,Brown:1986nw} have been
performed in the Hamiltonian formalism, the brackets of the surface
charges have been evaluated indirectly through \eqref{eq:5}. Indeed, a
direct computation in terms of fundamental canonical variables is much
more involved and is achieved, for instance, through the explicit
construction of the dual boundary theory and its degrees of freedom,
viz., Liouville theory \cite{Coussaert:1995zp} in the asymptotically
${\rm AdS}_3$ case. 

\section{Standard examples in three dimensions}
\label{sec:stand-exampl-three}

\subsection{Solution space and transformation laws}
\label{sec:solt-space-transf}

On-shell, three dimensional spacetimes that are asymptotically AdS at
spatial infinity or flat at null infinity are described by metrics of
the form
\begin{equation}
  \label{eq:6}
    ds^2=\left(-\frac{r^2}{l^2}+\mathcal{M}\right)du^2-2du
  dr+2\mathcal{N}du d\phi+r^2d\phi^2,
\end{equation}
where
\begin{equation}
  \label{eq:7}
   \mathcal{M}=2(\Xi_{++}+\Xi_{--}),\quad 
\mathcal{N}=l(\Xi_{++}-\Xi_{--}),
\end{equation}
with $\Xi_{\pm\pm}=\Xi_{\pm\pm}(x^\pm)$, $x^\pm= \frac{u}{l}\pm \phi$,
in the ${\rm AdS}$ case while $l\to\infty$ and
\begin{equation}
  \label{eq:8}
   \mathcal{M}=\Theta,\quad
  \mathcal{N}=\Xi+\frac{u}{2}\d_\phi \Theta,
\end{equation}
with $\Theta=\Theta(\phi)$ and $\Xi=\Xi(\phi)$ in the flat case.

In the ${\rm AdS}$ case, the easiest solutions where these functions
  are constants,
\begin{equation}
  \Xi_{\pm\pm}=2G(M\pm \frac{J}{l}),\label{BTZ}
\end{equation}
include both the BTZ black holes for which $M\geq 0$, $|J|\leq Ml$ and
the ${\rm AdS}_3$ spacetime which corresponds to $M=-\frac{1}{8G}$,
$J=0$. In the flat case, the zero mode solutions are given by
$\Theta=8GM$, $\Xi=4GJ$ and correspond for $M\geq 0$ to cosmological
solutions.

In the former case, the asymptotic symmetry algebra forms a
three-dimensional representation of the two dimensional conformal
algebra, described in terms of vector fields $\xi=Y^+\d_++ Y^-\d_-$,
$Y^\pm=Y^\pm(x^\pm)$, equipped with the standard Lie bracket, while in
the latter case, it is a representation of the $\mathfrak{bms}_3$
algebra described by vector fields $\xi=Y\d_\phi+(T+uY')\d_u$,
$Y=Y(\phi),T=T(\phi)$. 

By using that asymptotic symmetries preserve solutions, the
gravitational computation yields the transformation laws
\begin{equation}
  \label{eq:9}
  -\delta_\xi \Xi_{\pm\pm}=Y^\pm \Xi'_{\pm\pm}+2 Y^{\pm\prime}
  \Xi_{\pm\pm}-\half Y^{\pm\prime\prime\prime},
\end{equation}
respectively
\begin{equation}
  \label{eq:10}
 \begin{split}
   -\delta_\xi\, \Theta & = Y \Theta'+2 Y' \Theta -
   2  Y^{\prime\prime\prime}, \\
   -\delta_\xi\, \Xi & = Y \Xi'+ 2 Y' \Xi  +\half
   T\Theta'+  T' \Theta-
   T^{\prime\prime\prime}. 
 \end{split}
\end{equation}

\subsection{Current algebra}
\label{sec:current-algebra}

When the main purpose is to find the integrated surface charges
$Q_\xi=\int^{2\pi}_0 d\phi J^u_\xi$, what is usually explicitly worked out
from \eqref{eq:12} is the $k^{ur}_\xi$ and also $J^u_\xi$ component
only. In the cases at hand, when the background solution is fixed to
be the BTZ black hole with $M=0=J$ or the null orbifold respectively,
these components are given by 
\begin{equation}
  \label{eq:3}
\begin{split}
  J^u_\xi&= \frac{l}{8\pi
    G}[Y^+\Xi_{++}+Y^-\Xi_{--}], \\
 J^u_\xi&= \frac{1}{16\pi
    G}[T\Theta+2 Y\Xi]. 
\end{split}
\end{equation}

Instead of a direct computation of the spatial components, they can
easily be worked out from current conservation $\d_a J^a_f\approx 0$. 
This gives 
\begin{equation}
  \label{eq:4}
    J^\pm_\xi =\frac{1}{4\pi G} Y^\mp\Xi_{\mp\mp},
\end{equation}
in the ADS case, while $J^\phi_\xi=0$ in the flat case. 

The current algebra can then be deduced directly from the expression
for the currents, the asymptotic symmetry algebra and the
transformation laws of the fields,
\begin{equation}
  \label{eq:13}
  -\delta_{\xi_2} J^a_{\xi_1}\approx J^a_{[\xi_1,\xi_2]}+ K^a_{\xi_1,\xi_2}
  +\d_b L_{\xi_1,\xi_2}^{[ab]}, 
\end{equation}
where for ${\rm AdS}_3$, 
\begin{equation}
  \label{eq:14}
\begin{split}
  K^\pm_{\xi_1,\xi_2}&= \frac{1}{16\pi G}[ Y^{\mp\prime}_1
  Y^{\mp\prime\prime}_2-(1\leftrightarrow 2)], \\
  L_{\xi_1,\xi_2}^{[\pm\mp]}&=\frac{1}{4\pi G}[Y^\mp_1 Y^\mp_2
  \Xi_{\mp\mp} -\half  Y^{\mp}_1 Y^{\mp\prime\prime}_2 +\frac{1}{4} 
  Y^{\mp\prime}_1 Y^{\mp\prime}_2], 
\end{split}
\end{equation}
while 
\begin{equation}
  \label{eq:15}
  \begin{split}
    K^u_{\xi_1,\xi_2}&=\frac{1}{16\pi
      G}[Y_1'T_2''+T_1'Y_2''-(1\leftrightarrow 2)], \quad K^\phi_{\xi_1,\xi_2}=0,\\
L^{[u\phi]}_{\xi_1,\xi_2}&= \frac{1}{16\pi G}[ (Y_1 T_2 +
 T_1 Y_2) \Theta + 2( Y_1 Y_2 \Xi - T_1 Y''_2 -Y_1 T''_2) +T_1'Y_2'+Y_1'T_2'], 
  \end{split}
\end{equation}
in the flat case.

\section{Four dimensional asymptotically flat spacetimes at $\scri^+$}
\label{sec:four-dimens-asympt}

\subsection{Solution space and transformation laws}
\label{sec:solut-space-transf}

Our conventions are as in \cite{Barnich:2011ty}. More details can be
found for instance in the reviews
\cite{newman:1980xx,Penrose:1984,Penrose:1986,stewart:1991}. 

On the space-like cut of $\scri^+$, we use coordinates
$\zeta,\xbar\zeta$, the metric $d\xbar s^2=\xbar\gamma_{AB}
dx^Adx^B=2P^{-2}d\zeta d\xbar\zeta$ and the volume
element\footnote{There is a mistake of a factor 2 for
  $d^2\Omega^\varphi$ in \cite{Barnich:2011ty} after equation
  (6.5). For simplicity of notations, we have also removed the bar
  over the real scalar curvature $R$ of the metric $d\xbar s^2$.  },
$d^2\Omega^\varphi=(id\xbar\zeta\wedge d\zeta) P^{-2}$. For the unit
sphere, we have $\zeta=\cot{\frac{\theta}{2}}e^{i\phi}$ in terms of
standard spherical coordinates and $P_{S}=\frac{1}{\sqrt
  2}(1+\zeta\xbar\zeta)$ so that $d^2\Omega_{S}=\sin\theta
d\theta\wedge d\phi=\frac{2 d\zeta\wedge d\xbar\zeta}{i
  (1+\zeta\xbar\zeta)^2}$.

The covariant derivative on the $2$ surface is then encoded in the
operator
\begin{equation}
\eth \eta^s= P^{1-s}\xbar \d(P^s\eta^s),\qquad \xbar \eth
\eta^s=P^{1+s}\d(P^{-s}\eta^s)\,, 
\label{eq:34}
\end{equation}
where
$\eth,\xbar\eth$ raise respectively lower the spin weight by one
unit and satisfy  
\begin{equation} 
[\xbar \eth, \eth]\eta^s=\frac{s}{2}  R\,
  \eta^s\,,\label{eq:35}
\end{equation}
with $ R=4P^2\d \xbar \d \ln P$, $ R_S=2$. The spin weights of
the various quantities are summarized in table \ref{t1}. Note that a
field $\eta$ of spin weight $s$ and conformal weight $w$ transforms as
\begin{equation}
  \label{eq:52}
  \delta_{\cY,\xbar \cY}\eta=\big[\cY \eth + \xbar \cY \xbar \eth + h
  \eth \cY +\xbar h \xbar \eth \xbar \cY \big] \eta,
\end{equation}
where the conformal dimensions are given by $(h,\xbar
h)=(\frac{s-w}{2},\frac{-s-w}{2})$. 

Let $\cY=P^{-1} \xbar Y$ and $\xbar \cY=P^{-1}
Y$. The conformal Killing equations and the conformal factor then
become
\begin{equation}
  \label{eq:25}
  \eth \xbar \cY=0=\xbar\eth \cY,\qquad \psi=(\eth \cY+\xbar\eth
  \xbar \cY)\,.
\end{equation}
It follows for instance that 
\begin{equation}
\xbar\eth \eth \cY=-\frac{
  R}{2}\cY,\quad \eth^2 \psi=\eth^3\cY-\frac{1}{2}\xbar\cY\eth  R,\quad 
\xbar\eth\eth \psi=-\frac{1}{2}[\eth( R\cY)+\xbar\eth(
R\xbar\cY)]\label{eq:44}\,.
\end{equation}

Let $f=T+\half u\psi$ and $\xi=f\d_u+\cY\eth+\xbar\cY\xbar \eth$ with
$\cY,\xbar\cY$ conformal Killing vectors. The asymptotic symmetry
algebra is described by the vector fields $\xi$ on $\scri^+$,
parametrized by $T,\cY,\xbar\cY$, and equipped with the standard Lie
bracket. More explicitly, $\hat \xi=[\xi_1,\xi_2]$ is parametrized by
$\hat T, \hat\cY, \hat {\xbar\cY}$ where
\begin{equation}
\begin{gathered}
  \label{eq:57}
  \hat \cY=\cY_1\eth \cY_2 -(1\leftrightarrow 2),\qquad  \hat
  {\xbar\cY}=\xbar \cY_1\xbar \eth \xbar \cY_2 -(1\leftrightarrow 2),\\
  \hat T= (\cY_1\eth +\xbar \cY_1\xbar \eth)T_2-\half \psi_1 T_2 -(1\leftrightarrow 2)\,.
\end{gathered}
\end{equation}

The part of solution space that is relevant for the asymptotic current
algebra on $\scri^+$ is given by fields $\sigma^0,\Psi^0_2,\Psi^0_1$
and their complex conjugates and are denoted collectively by
$\chi$. In this framework, $\dot\sigma^0$ is the news function. For
convenience, one also introduces
\begin{equation}
  \label{eq:53}
  \Psi^0_{3}=-\eth
\dot{\xbar\sigma}^0-\frac{1}{4}\xbar\eth R, \qquad \Psi^0_{4}=-\ddot{\xbar\sigma}^0\,.
\end{equation}
In these terms, the evolution equations are 
\begin{equation}
  \label{eq:41}
\dot \Psi^0_3=\eth \Psi^0_4,\quad 
\dot \Psi^0_{2}=\eth
\Psi^0_{3}+ \sigma^0\Psi^0_{4},\qquad  \dot \Psi^0_{1}=\eth
\Psi^0_{2}+2 \sigma^0\Psi^0_{3}\,,
\end{equation}
which have to be supplemented by the additional on-shell relation,
\begin{equation}
  \label{eq:48}
  \Psi^0_2-\xbar\Psi^0_2=\xbar\eth^2\sigma^0-\eth^2\xbar\sigma^0
+\dot \sigma^0\xbar\sigma^0-\sigma^0\dot{\xbar\sigma}^0\,,
\end{equation}
and the transformation laws are
\begin{equation}
\begin{split}
  \label{eq:16b}
  -\delta_\xi \sigma^0 & = [f\d_u+\cY\eth+ \xbar
  \cY\xbar\eth+\frac{3}{2}\eth \cY-\frac{1}{2} \xbar\eth \xbar \cY] \sigma^0-\eth^2
  f\,,\\
  -\delta_\xi \dot\sigma^0 & = [f\d_u+ \cY\eth + \xbar \cY\xbar\eth+2\eth
  \cY]\dot\sigma^0-\half \eth^2 \psi\,,\\
-\delta_\xi\Psi^0_4&=[f\d_u+\cY\eth+\xbar\cY\xbar\eth+\half \eth\cY
+\frac{5}{2}\xbar\eth\xbar\cY]\Psi^0_4\,,\\
-\delta_\xi\Psi^0_3&=[f\d_u+\cY\eth+\xbar\cY\xbar\eth+\eth\cY
+2\xbar\eth\xbar\cY]\Psi^0_3+\eth f\Psi^0_4\,,\\
  -\delta_\xi
  \Psi^0_2&=[f\d_u+\cY\eth+\xbar\cY\xbar\eth+\frac{3}{2}\eth \cY
  +\frac{3}{2}\xbar\eth \xbar\cY ]\Psi^0_2
  +2\eth f\Psi^0_3,\\
  -\delta_\xi \Psi^0_1&
  =[f\d_u+\cY\eth+\xbar\cY\xbar\eth+2\eth\cY+\xbar\eth\xbar\cY
  ]\Psi^0_1  +3\eth f\Psi^0_2\,.
\end{split}
\end{equation}

\begin{table} 
\caption{Spin and conformal weights}\label{t1}
\begin{center}
\begin{tabular}{c|c|c|c|c|c|c|c|c} & $\sigma^0$  & $\dot\sigma^0$ &
  $\Psi^0_4 $&  $\Psi^0_3$ & $\Psi^0_2$ & $\Psi^0_1$ & $\cY$ & $T$  \\
\hline
s &  $2$ &  $2$  & $-2$  & $-1$ & $0$ & $1$ &  $-1$ & $0$ \\ 
\hline
w  & $- 1$  & $-2$  & $-3$ & $-3$ & $-3$ & $-3$ & $1$ & $1$ \\  
\end{tabular}
\end{center} \end{table}

\subsection{Current algebra in the absence of news}
\label{sec:curr-algebra-absence}

Let us concentrate on the sphere $P=P_S$, or the Riemann sphere with
$P=P_R=1$, so that $\eth  R=0=\xbar\eth  R$. Let us also assume
that there is no news, $\dover{^n}{u^n}\sigma=0$ for all $n\geq 1$,
which implies in particular also that $\Psi^0_4=0=\Psi^0_3$ and that
their time derivatives vanish. Furthermore, the complex conjugates of
all these expressions also vanish. The transformations that leave
these conditions invariant have to satisfy $\eth^2\psi=0=\xbar\eth^2
\psi$. Both on the Riemann and the unit sphere, this is equivalent to
$\eth^3\cY=0=\xbar\eth^3\xbar\cY$ and then to $\cY=P^{-1}\xbar Y$,
$\xbar\cY =P^{-1} Y$ with $\xbar \d^3 \xbar Y=0=\d^3 Y$. It follows that
among the superrotations, only the standard Lorentz transformations
corresponding to $l_{-1}=P^{-1}\eth$, $l_0=P^{-1}\xbar\zeta\eth$,
$l_1=P^{-1}\xbar \zeta^2\eth$ and their complex conjugates remain.  The
remaining dynamical equations simplify and read $\dot \Psi^0_2=0$,
$\dot \Psi^0_1=\eth \Psi^0_2$, together with their complex conjugates,
and also $\Psi^0_2-\xbar\Psi^0_2= \xbar\eth^2\sigma^0
-\eth^2\xbar\sigma^0$. In the following, absence of news means that
all the above conditions are satisfied. 

Note that one could also impose $\eth^2 T=0=\xbar\eth^2 T$, in which
case the supertranslations reduce to ordinary translations and the
asymptotic symmetry algebra becomes the Poincar\'e algebra. We will
not make these additional assumptions here.

In the absence of news, the time component of the current, which is
real, is given by \eqref{eq:21} where the second term proportional to
$\dot{\xbar\sigma}^0$ is absent. Current conservation in the form
\begin{equation}
\d_u \cJ^u_\xi+\eth \cJ_\xi +\xbar\eth \xbar{\cJ_\xi}\approx 0,\label{eq:23}
\end{equation}
then leads to $\cJ_\xi$ given in \eqref{eq:36}, where the second, the
third and the last two terms vanish according to the assumptions made
in the present section. This is equivalent to the standard
conservation law
\begin{equation}
\d_a J^a_\xi\approx 0,\label{eq:17}
\end{equation}
provided that $J^u_\xi=P^{-2}
\cJ^u_\xi$, $J_\xi^{\xbar\zeta}=P^{-1} \cJ_\xi$,
$J_\xi^{\zeta}=P^{-1}\xbar{\cJ_\xi}$.

Current algebra in the absence of news is then explicitly given by
\begin{equation}
  \label{eq:22}
  \begin{split}
& -\delta_{\xi_2} \cJ^u_{\xi_1}\approx \cJ^u_{[\xi_1,\xi_2]}+
\eth \cL_{\xi_1,\xi_2}+\xbar\eth
\xbar{\cL_{\xi_1,\xi_2}},\\
& -\delta_{\xi_2} \cJ_{\xi_1}\approx \cJ_{[\xi_1,\xi_2]}-\d_u
\cL_{\xi_1,\xi_2}+\xbar\eth \xbar{\cM_{\xi_1,\xi_2}}, 
  \end{split}
\end{equation}
where $\cL_{\xi_1,\xi_2}$ is obtained from \eqref{eq:23a} and
$\xbar{\cM_{\xi_1,\xi_2}}=-\cM_{\xi_1,\xi_2}$ from \eqref{eq:16} by
dropping all terms that vanish in the absence of news. It is
understood that the algebra for $\xbar{\cJ_\xi}$ is obtained by the
one for $\cJ_\xi$ through complex conjugation. This is equivalent to
the standard form
\begin{equation}
-\delta_{\xi_2}
J^a_{\xi_1}\approx J^a_{[\xi_1,\xi_2]}+\d_b L^{[ab]}_{\xi_1,\xi_2},\label{eq:18}
\end{equation}
provided that $L^{[u\xbar\zeta]}_{\xi_1,\xi_2}=P^{-1}
\cL_{\xi_1,\xi_2}$,
$L^{[u\zeta]}_{\xi_1,\xi_2}=P^{-1}\xbar{\cL_{\xi_1,\xi_2}}$,
$L^{[\xbar\zeta\zeta]}_{\xi_1,\xi_2}= \xbar{\cM_{\xi_1,\xi_2}}$. In
particular, this requires $\cM_{\xi_1,\xi_2}$ to be purely imaginary.

Consistency conditions between conservation and current algebra can be
deduced by using that $\d_u \cJ^u_\xi=\dover{}{u}
\cJ^u_\xi-\delta_{\d_u} \cJ^u_\xi$, where the partial derivative denotes
the explicit $u$ dependence contained in $\xi$, and also
$\cJ^u_{[\xi,\d_u]}=\cJ^u_{-\d_u\xi}$. It then follows from \eqref{eq:23} and
\eqref{eq:22} that
\begin{equation}
  \label{eq:30}
  \eth (\cJ_\xi +\cL_{\xi,\d_u})+\xbar\eth (\xbar{ \cJ_\xi}
  +\xbar{\cL_{\xi,\d_u}})\approx 0. 
\end{equation}
This condition is satisfied because \eqref{eq:23a} implies that
$\cJ_\xi=-\cL_{\xi,\d_u}$.

\subsection{Current algebra in the presence of news}
\label{sec:curr-algebra-pres}

As in standard applications (see e.g. \cite{Treiman:1985}), current
algebra is more involved when currents are not conserved. This is the
case in the presence of news. We will also allow for an arbitrary,
$u$-independent, $P(\zeta,\xbar\zeta)$ in the metric on the space-like
cut of $\scri^+$. The presence of news implies in addition that
$k^{[ur]}_\xi,k^{[Ar]}_\xi,$ are no longer integrable.

Let 
\begin{equation}
  \label{eq:26}
  \Theta^u_\xi(\delta\chi)=\frac{1}{8\pi G}
 \Big[ f\dot{\xbar\sigma}^0\delta\sigma^0+{\rm c.c.}\Big], 
\end{equation}
and 
\begin{equation}
  \label{eq:27}
  \cK^u_{\xi_1,\xi_2}=\frac{1}{8\pi G} \Big[\big( 
  \half \xbar \sigma^0f_1 \eth^2 \psi_2+
  \frac{1}{4} f_1\eth
  f_2\xbar\eth  R
  -(1\leftrightarrow 2) \big)+{\rm c.c.}\Big],
\end{equation}
be expressions that vanish in the absence of news. The computations
of \cite{Barnich:2011mi}, translated to the Newman-Penrose formalism
in \cite{Barnich:2011ty}, state that the time component of the current
\begin{equation}
  \label{eq:21}
   \cJ^u_\xi=-\frac{1}{8\pi G}\Big[\big(
  f(\Psi^0_2+\sigma^0\dot{\xbar\sigma}^0)+\cY(\Psi^0_1
  +\sigma^0\eth\xbar\sigma^0+\half\eth(\sigma^0\xbar\sigma^0))\big)
  +{\rm c.c.}\Big]. 
\end{equation}
satisfies 
\begin{equation}
  \label{eq:28}
  -\delta_{\xi_2}
  \cJ^u_{\xi_1}+\Theta^u_{\xi_2}(-\delta_{\xi_1}\chi)\approx\cJ^u_{[\xi_1,\xi_2]}+
\cK^u_{\xi_1,\xi_2}+
\eth\cL_{\xi_1,\xi_2} +\xbar\eth\xbar{\cL_{\xi_1,\xi_2}},
\end{equation}
for some $\cL_{\xi_1,\xi_2}$. Since it is relevant for current
algebra, the explicit expression is now worked out. It is given by
\begin{multline}
  \cL_{\xi_1,\xi_2} = \cY_2 \cJ^u_{\xi_1}-f_2\cJ_{\xi_1}\\-\frac{1}{8\pi
    G}\Big[ (\half[ \eth\cY_1+\xbar\eth\xbar\cY_1]\eth f_2-\half
  \cY_1\eth^2 f_2 -\xbar\cY_1 \eth\xbar\eth f_2 ) \xbar\sigma^0\\
  -\frac{1}{2} \cY_1\xbar\eth^2 f_2 \sigma^0 +(-\cY_1\eth f_2+
  \xbar\cY_1\xbar\eth f_2)\eth\xbar\sigma^0 - f_1 \eth f_2
  \dot{\xbar\sigma}^0 \Big].\label{eq:23a}
\end{multline}
where  
\begin{multline} 
\cJ_\xi=\frac{1}{8\pi G}\Big[\cY\big(\Psi^0_2 +\half [\dot
\sigma^0\xbar\sigma^0-\sigma^0 \dot{\xbar\sigma}^0] \big)-\half
\eth (\eth\cY -\xbar\eth \xbar\cY) \xbar\sigma^0
\\+\half(\eth\cY-\xbar\eth\xbar\cY) \eth\xbar\sigma^0
+f\Psi^0_3+\eth f\dot{\xbar\sigma}^0
\Big].\label{eq:36}
\end{multline}
In order to identify the spatial component of the current
$\cJ_{\xi}$, we have used that the non-conservation of $J^u_\xi$ can
be obtained from \eqref{eq:28} for $\xi_1=\xi$ and
$\xi_2=\d_u$. Indeed, this gives
\begin{equation}
  \label{eq:25a}
  \d_u \cJ^u_\xi+\eth \cJ_\xi+ \xbar\eth\xbar{\cJ_\xi}
\approx \Theta^u_{\d_u}(\delta_\xi \chi)+\cK^u_{\xi,\d_u}, 
\end{equation} 
when $\cJ_\xi=-\cL_{\xi,\d_u}$. 

Alternatively, one can obtain the spatial components directly by
evaluating the $k^{[Ar]}$-components of the surface charge $n-2$
form. This is done explicitly in the appendix. It also provides an
expression for the non-integrable part,
\begin{equation}
  \label{eq:19}
  \Theta_\xi [\delta \chi] = \frac{1}{8 \pi G}\, \cY\Big[  \dot
  {\xbar \sigma}^0 \delta \sigma^0 + \dot \sigma^0 \delta \xbar
  \sigma^0 \Big]
=\cY\Theta^u_{\d_u}(\delta\chi).  
\end{equation}

One can now work out the current algebra for the spatial component, 
\begin{equation}
  \label{eq:11}
  -\delta_{\xi_2}
  \cJ_{\xi_1}+\Theta_{\xi_2}(-\delta_{\xi_1}\chi)\approx \cJ_{[\xi_1,\xi_2]}+
  \cK_{\xi_1,\xi_2}-\d_u\cL_{\xi_1,\xi_2} +\xbar\eth\xbar{  \cM_{\xi_1,\xi_2}},
\end{equation}
where 
\begin{multline}
  \label{eq:20}
  \cK_{\xi_1,\xi_2}
=\frac{1}{8\pi G}\Big[
\half\sigma^0\cY_1\xbar\eth^2\psi_2+\half\xbar\sigma^0\cY_1
\eth^2\psi_2+\half \xbar\eth^2 \psi_1\eth f_2\\
 +\frac{1}{4} \eth R \cY_1\xbar\eth f_2 +\frac{1}{4} \xbar \eth R
 \cY_1\eth f_2-(1\leftrightarrow 2)\Big],
\end{multline}
and 
\begin{equation}
  \label{eq:16}
  \xbar{\cM_{\xi_1,\xi_2}}=\Big(\xbar\cY_2 \cJ_{\xi_1}+\frac{1}{8\pi
    G}\Big[\half\eth (\eth \cY_1-\xbar\eth \bar\cY_1) \xbar\eth f_2-\half\eth
\cY_1\eth\xbar\eth f_2 
\Big]-{\rm c.c.}\Big). 
\end{equation} 
If one defines in addition
$\theta^u_\xi=P^{-2}\Theta^u_\xi$,
$\theta^{\xbar\zeta}_\xi=P^{-1}\Theta_\xi$,
$\theta^{\zeta}_\xi=P^{-1}\xbar{\Theta_\xi}$ and also
$K^u_{\xi_1,\xi_2}=P^{-2} \cK^u_{\xi_1,\xi_2}$,
$K_{\xi_1,\xi_2}^{\xbar\zeta}=P^{-1} \cK_{\xi_1,\xi_2}$,
$K_{\xi_1,\xi_2}^{\zeta}=P^{-1}\xbar{\cK_{\xi_1,\xi_2}}$, the total
current algebra takes the form
\begin{equation}
  \label{eq:33}
  -\delta_{\xi_2} J^a_{\xi_1}+\theta^a_{\xi_2}(-\delta_{\xi_1}\chi)\approx J^a_{[\xi_1,\xi_2]}+
  K^a_{\xi_1,\xi_2}+\d_b L^{[ab]}_{\xi_1\xi_2}. 
\end{equation}

The time component of the field-dependent central term has
been shown to satisfy the cocycle condition
\begin{equation}
  \label{eq:29}
  \cK^u_{[\xi_1,\xi_2],\xi_3}-\delta_{\xi_3} \cK^u_{\xi_1,\xi_2}+{\rm
    cyclic} (1,2,3)=\eth \cN_{\xi_1,\xi_2,\xi_3}+\xbar\eth
  \xbar{\cN_{\xi_1,\xi_2,\xi_3}}, 
\end{equation}
for some $\cN_{\xi_1,\xi_2,\xi_3}$, which is now worked out to be
\begin{equation}
 \cN_{\xi_1,\xi_2,\xi_3} =- f_3 \cK_{\xi_1,\xi_2} +{\rm cyclic}
 (1,2,3). 
\end{equation}
This is completed by showing that the spatial component satisfies 
\begin{equation}
  \label{eq:31}
  \cK_{[\xi_1,\xi_2],\xi_3}-\delta_{\xi_3} \cK_{\xi_1,\xi_2}+{\rm
    cyclic} (1,2,3)= -\d_u \cN_{\xi_1,\xi_2,\xi_3}+\xbar\eth
  \xbar{\cO_{\xi_1,\xi_2,\xi_3}}, 
\end{equation}
where
\begin{multline}
  \label{eq:32}
  \xbar{\cO_{\xi_1,\xi_2,\xi_3}}=\Big(\frac{1}{8\pi G} \xbar{\cY_3}
  \Big[\half\sigma^0(\cY_1\xbar\eth^3\xbar{\cY_2}
  -\cY_2\xbar\eth^3\xbar{\cY_1}) +\half (\xbar\eth^3\xbar{\cY_1}\eth
  f_2 -\xbar\eth^3\xbar{\cY_2}\eth f_1)\Big]\\ -{\rm c.c.}\Big) +{\rm
    cyclic} (1,2,3).
\end{multline}
If one defines $N^{[u\xbar\zeta]}_{\xi_1,\xi_2,\xi_3}=P^{-1}
\cN_{\xi_1,\xi_2,\xi_3}$,
$N^{[u\zeta]}_{\xi_1,\xi_2,\xi_3}=P^{-1}\xbar{\cN_{\xi_1,\xi_2,\xi_3}}$,
$N^{[\xbar\zeta\zeta]}_{\xi_1,\xi_2,\xi_3}=
\xbar{\cO_{\xi_1,\xi_2,\xi_3}}$, the complete field dependent
extension is thus a current of the dual boundary theory that satisfies
\begin{equation}
K^a_{[\xi_1,\xi_2],\xi_3}-\delta_{\xi_3} K^a_{\xi_1,\xi_2}+{\rm
    cyclic} (1,2,3)=\d_b N^{[ab]}_{\xi_1,\xi_2,\xi_3}.
\end{equation}

The current algebra \eqref{eq:33} is valid both in the case of
$\mathfrak{bms}_4^{\rm loc}$, in which case one may choose for example
to expand $Y,\bar Y,PT$ in Laurent series, and for
$\mathfrak{bms}_4^{\rm glob}$, which is explicitly obtained by using
$P=P_S$ and restricting oneself to Lorentz transformations as
described in the beginning of section \ref{sec:curr-algebra-absence},
while simultaneously expanding $T$ in spherical harmonics. In this
case, all components of the extension $K^a_{\xi_1,\xi_2}$ are easily
seen to vanish.

\section*{Acknowledgements}
\label{sec:acknowledgements}

This work is supported in part by the Fund for Scientific
Research-FNRS (Belgium), by IISN-Belgium, by ``Communaut\'e fran\c
caise de Belgique - Actions de Recherche Concert\'ees''. The Centro de
Estudios Cient\'ificos (CECs) is funded by the Chilean Government
through the Centers of Excellence Base Financing Program of Conicyt.

\appendix

\section{Direct derivation of spatial components}
\label{sec:direct-deriv-spat}

The spatial components of the currents are computed in two steps. We
will start with the conventions of \cite{Barnich:2010eb} and then we
will use the dictionary given in \cite{Barnich:2011ty} to translate
the result into the Newman-Penrose formalism used in section
\ref{sec:four-dimens-asympt}. The non-integrable piece is identified
as in \cite{Wald:1999wa} and treated in the current algebra as in
\cite{Barnich:2011mi}.

Asymptotically flat metrics solving Einstein's equation are of the form
\begin{equation}
\label{eq:assymptmetricI}
ds^2 = e^{2 \beta} \frac{V}{r} du^2 - 2 e^{2\beta} du dr + g_{AB}(dx^A
- U^A du)(dx^B - U^B du),
\end{equation}
where
\begin{equation}
g_{AB} = r^2 \xbar \gamma_{AB} + r C_{AB} + \frac{1}{4} \xbar
\gamma_{AB} C^C_D C^D_C + o(r^{-\epsilon}). 
\end{equation}
The 2D metric $\xbar \gamma_{AB}$ is fixed and corresponds to a choice
of $P$. Indices on $C_{AB}$ are raised with the inverse of $\xbar
\gamma_{AB}$ and $C^A_A = 0$. To the order we need, the other metric
components are given by
\begin{gather}
  \beta = -\frac{1}{32}r^{-2} C^A_B C^B_A + o(r^{-3-\epsilon}),\\
  g_{uA} = \frac{1}{2} \xbar D_B C^B_A + \frac{2}{3} r^{-1} \left[N_A
    + \frac{1}{4} C_{AB}\xbar D_C C^{CB}\right] +
  o(r^{-1-\epsilon}),\\
  \frac{V}{r} = -\frac{1}{2} R + r^{-1} 2 M +
  o(r^{-1-\epsilon}),\label{eq:assymptmetreicV}
\end{gather}
where $\xbar D_A$ is the covariant derivative associated to
$\xbar\gamma_{AB}$ and $R$ is its scalar curvature. In these
conventions, $M(u, x^A)$, $N_A(u, x^A)$ and $C_{AB}(u, x^A)$
parametrize the part of the solution space relevant for the asymptotic
current algebra. They will be related below to the fields $\sigma^0$,
$\Psi^0_1$ and $\Psi_2^0$. 

The ${\rm BMS}_4$ algebra parametrized by $\xi = f \d_u + Y^A \d_A$,
$f=T + \frac{1}{2} u \xbar D_A Y^A$, is represented by space-time
vector fields as follows:
\begin{equation}
\left\{\begin{array}{l} {}^{(4)}\xi^u = f,\\ {}^{(4)}\xi^A = Y^A +
    I^A, 
\qquad I^A =
    -\d_B f \int_r^\infty dr'(e^{2\beta} g^{AB}),\\ {}^{(4)}\xi^r =
    -\frac{1}{2} r (\xbar D_A {}^{(4)}\xi^A - \d_A f
    U^A).\end{array}\right. 
\end{equation}
The associated $n-2$ form $k^{[\mu\nu]}_\xi$ is then computed using
the expression given in \cite{Barnich:2001jy}, 
\begin{multline}
  \label{eq:29a}
k^{[\mu\nu]}_\xi[h, g]= \frac{1}{16
    \pi G}\sqrt{- g}\Big[{}^{(4)}\xi^\nu D^\mu h
  -{}^{(4)}\xi^\nu D_\sigma h^{\mu\sigma} +{}^{(4)}\xi_\sigma D^\nu
  h^{\mu\sigma}
  \\
  +\frac{1}{2}h D^\nu {}^{(4)}\xi^\mu +\half
  h^{\nu\sigma}(D^\mu {}^{(4)}\xi_\sigma-D_\sigma {}^{(4)}\xi^\mu)
  -(\mu\leftrightarrow \nu)\Big]\,,
\end{multline}
where $h_{\mu\nu}$ stands for a variation of the metric $\delta
g_{\mu\nu}$ preserving the asymptotic structure. The information
relevant for the asymptotic current algebra is given by the components
$k^{[ur]}_\xi$ and $k^{[Ar]}_\xi$ evaluated in the limit $r
\rightarrow \infty$. The former has already been computed in
\cite{Barnich:2011mi},
\begin{multline}
  k_\xi^{[ur]}= \frac{\sqrt{\xbar\gamma}}{16 \pi G} \lim_{r
    \rightarrow \infty}\Big[ rY^A \xbar D_B \delta C^B_A -
  \frac{1}{8}\xbar D_C Y^C C^{AB} \delta C_{AB}+4 f \delta M
  \\-\frac{1}{2} f \xbar D_A \xbar D_B \delta C^{AB} + \frac{1}{2} f
  \delta C^{AB}\d_u C_{AB} -\frac{1}{2} \xbar D_A f \xbar D_B \delta
  C^{AB}+2 Y^A \delta N_A\big].
\end{multline}
The latter are now shown to be given by 
\begin{multline}
  k_\xi^{[Ar]}= \frac{\sqrt{\xbar\gamma}}{16 \pi G} \lim_{r
    \rightarrow \infty}\Big[-r Y^B\d_u \delta C^A_B - 2Y^A \delta M
  -\frac{1}{2} \delta C^A_B \xbar D^B \xbar D_C Y^C +\xbar D_Bf\d_u
  \delta C^{BA} \\ - \frac{1}{2} f \xbar D_B\d_u \delta C^{BA} -\frac{
    R}{2}\delta C^A_BY^B - \frac{1}{2} Y^A \xbar D_B\xbar D_C \delta
  C^{BC} +\frac{1}{2} Y^B \xbar D^A\xbar D_C \delta C^C_B \\ -
  \frac{5}{8} Y^A C^{BC}\d_u \delta C_{BC} + Y^B C^{CA}\d_u \delta
  C_{BC} -\frac{1}{8} Y^A \delta C^{BC}\d_u C_{BC} \\ + Y^B \delta
  C^{CA}\d_u C_{BC} +\frac{1}{4} (\xbar D^B Y^A -\xbar D^A Y^B + 2
  \xbar D_D Y^D \xbar \gamma^{AB}) \xbar D_C\delta C^C_B \Big].
\end{multline}

In order to translate these results into the Newman-Penrose formalism,
we use the dictionary \cite{Barnich:2011ty},
\begin{gather}
C^{\xbar\zeta}_\zeta = 2 \xbar \sigma^0, \qquad C^\zeta_{\xbar \zeta} = 2
\sigma^0, \qquad\cY = P^{-1} Y^{\xbar \zeta}, \qquad \xbar\cY = P^{-1} Y^{ \zeta},\\
2M = - \Psi^0_2 - \xbar\Psi^0_2-\d_u(\sigma^0 \xbar \sigma^0),\\
P N_\zeta = -\xbar \Psi^0_1 - \xbar \sigma^0 \xbar \eth\sigma^0 -
\frac{3}{4}\xbar \eth (\sigma^0 \xbar \sigma^0), \qquad N_{\xbar\zeta}
= \xbar {N_\zeta},
\end{gather}
together with the definition of $\eth$ and $\xbar \eth$ given in
equation \eqref{eq:34}.  Straightforward computation then leads to 
\begin{multline}
  k^{[ur]}_\xi = \frac{P^{-2}}{8 \pi G} \lim_{r \rightarrow
    \infty}\Big[r \xbar \cY \eth \delta \xbar\sigma^0 - \frac{1}{2} f
  \eth^2 \delta \xbar \sigma^0 - \frac{1}{2} \eth f \eth \delta \xbar
  \sigma^0 - f \delta \Psi_2^0 - f \sigma^0 \delta \dot{\xbar
    \sigma}^0 \\ - \cY \delta \Psi_1^0 - \frac{1}{4}\cY \delta (7\sigma^0
  \eth \xbar \sigma^0 +3\eth \sigma^0 \xbar \sigma^0) - \frac{1}{4}
  \eth \cY (\sigma^0 \delta \xbar \sigma^0 + \xbar \sigma^0 \delta
  \sigma^0) + c.c.\Big],\label{eq:kurfull}
\end{multline}
\begin{multline}
  k^{[\xbar \zeta r]}_\xi = \frac{P^{-1}}{8 \pi G}\lim_{r \rightarrow
    \infty}\Big[- r \xbar\cY \delta \dot{\xbar \sigma} ^0 + \cY
  \delta\Psi^0_2 - \frac{1}{2}\cY \xbar \eth^2 \delta\sigma^0 +
  \frac{1}{2}\xbar \cY \,\xbar\eth\eth \delta\xbar\sigma^0 + \eth f
  \delta\dot{\xbar \sigma}^0 \\ -\frac{1}{2} f \eth \delta\dot {\xbar
    \sigma}^0 - \frac{1}{2} R \xbar \cY \delta\xbar \sigma^0 -
  \frac{1}{4} \cY \sigma^0 \delta\dot {\xbar \sigma}^0 + \frac{3}{4}
  \cY \xbar \sigma^0 \delta\dot\sigma^0 \\ +\frac{1}{4}\left( 3
    \eth\cY + \xbar\eth\xbar \cY \right)\eth \delta\xbar\sigma^0 -
  \frac{1}{2} \eth (\eth \cY + \xbar \eth \xbar \cY) \delta\xbar
  \sigma^0 + \frac{3}{4}\cY \delta \sigma^0 \dot {\xbar \sigma}^0 +
  \frac{7}{4}\cY \delta \xbar \sigma^0 \dot \sigma^0
  \Big],\label{eq:kzetarfull}
\end{multline}
and the last component $k^{[\zeta r]}_\xi$ being the complex conjugate
of the right hand side of \eqref{eq:kzetarfull}.

Expressions \eqref{eq:kurfull}-\eqref{eq:kzetarfull} diverge as $r$
goes to infinity. This is not a problem though as the divergent part
can be absorbed in an exact form
$\d_\rho\eta^{[\mu\nu\rho]}_\xi$. Defining $\eta_\xi^{[u r\xbar
  \zeta]}= P^{-1}\cN^u_\xi$, $\eta_\xi^{[u r \zeta]}=
P^{-1}\xbar{\cN^u_\xi}$ and $\eta_\xi^{[\xbar\zeta r
  \zeta]}=\xbar{\cN_\xi} $, we get 
\begin{gather}
  P^2 k_\xi^{[ur]} = \delta \cJ_\xi^u + \Theta_\xi^u[\delta \chi] +
  \eth \cN^u_\xi + \xbar {\eth \cN^u_\xi}, \\ P k_\xi^{[\xbar \zeta
    r]} = \delta \cJ_\xi + \Theta_\xi[\delta \chi] - \d_u \cN^u_\xi +
  \xbar {\eth \cN_\xi},
\end{gather}
where the currents $\cJ_\xi^u,\cJ_\xi$ are given in
\eqref{eq:21},\eqref{eq:36}, while the non-integrable pieces
$\Theta_\xi^u[\delta \chi],\Theta_\xi[\delta \chi]$ are given in
\eqref{eq:26}, \eqref{eq:19}, and
\begin{gather}
  \cN^u_\xi = \frac{1}{8 \pi G} \lim_{r \rightarrow \infty}\Big[r
  \xbar \cY \delta \xbar\sigma^0 - \frac{1}{2} f \eth \delta \xbar
  \sigma^0 - \frac{1}{4} \cY (\sigma^0 \delta \xbar \sigma^0 +
  \xbar \sigma^0 \delta \sigma^0) \Big],\\
  \xbar{\cN_\xi} = \frac{1}{8 \pi G}\Big[\frac{1}{2}\xbar \cY \eth
  \delta\xbar\sigma^0 - \frac{1}{2}\cY \xbar \eth \delta\sigma^0\Big].
\end{gather}

\def\cprime{$'$}
\providecommand{\href}[2]{#2}\begingroup\raggedright\endgroup

\bibliography{/Users/gbarnich/Documents/Literature/Bibliography/master2}

\end{document}